\documentstyle[12pt]{article}
\begin{document}
\centerline{\bf On the Aharonov--Anandan experiment}
\vskip 1cm
\centerline{\bf Carlo Rovelli}
\vskip.5cm
\centerline{Physics Dept. University of Pittsburgh,
Pittsburgh PA 15260, USA.}
\centerline{Dipart. di Fisica Universita' di Trento and INFN
sez. Padova,  Italia.}
\centerline{rovelli@pittvms.bitnet}

\vskip 2cm

\noindent{\bf Abstract}
\vskip .3cm

An experiment that would measure non--commuting
quantum mechanical observables without collapsing
the wave function has been recently proposed by
Y Aharonov and J Anandan.  These authors argue
that this "protected measurement" may give indication
on "the reality of the wave function".
We argue that, depending of the precise version of the experiment
considered, either the author's prediction is
incorrect and the wave function does collapse, or
the measurement is not a measurement on a quantum
system.  In either case, the experiment does not provide
a way for measuring non--commuting observables without
collapse, and it does not bear on the
issue of the "reality of the wave function".

\vskip 3cm

Yakir Aharonov and Jeeva Anandan have recently discussed the
problem of the interpretation of quantum mechanics, and have
proposed an experiment which could provide indications concerning
the "reality" of the wave function [1]. This proposal has received a
certain attention [2].  The experiment consists in a standard
Stern--Gerlach experiment, to which an additional homogeneous
strong
magnetic field $\vec B$ has been added.   According to these authors
the presence of this
additional magnetic field prevents the beam's trajectory
from splitting.  The authors then argue that one can reconstruct the
full initial wave function of a single particle (up to an overall
phase) by means of a
sequence of measurements of this kind on the same particle.  This
result seems
in contradiction with the generally accepted {\it credo\ }  that any
measurement disturbs the measured system, and would indeed
suggest a
more realistic view of the wave function than the one
commonly advocated.  In this note, we argue that this result is
either
incorrect, or unsubstantial.

We will not discuss the general theoretical framework of Aharonov
and
Anandan. Rather, we focus on the specific experiment proposed in
[1], namely the modified Stern-Gerlach experiment.
The main idea of the modified Stern--Gerlach experiment is to add a
strong homogeneous magnetic field to a conventional Stern-Gerlach
apparatus.  The authors claim that the effect of this additional
magnetic field is to prevent the beam from splitting (last
paragraph of pg. 9 of Ref.[1]).  Taken literally,
this claim is false. Indeed, if the
additional magnetic field $\vec B$ is in a (arbitrary, but)
fixed direction, while the beam has an arbitrary initial polarization,
then the beam will certainly split.\footnote{We consider here
spin 1/2 particles. By saying that the initial state $\Psi$ is
polarized in the direction  $\vec r$, we mean that $\Psi$ is the
positive-eigenvalue eigenstate of the the spin operator in the
direction $\vec r$, namely of the operator $\hat s_{\vec r}=\vec r
\cdot \vec\sigma$, where
$\vec\sigma$ are the Pauli matrices.}   More precisely, in the limit
in which the homogeneous $\vec B$ field is large,
the beam will split into the two eigenstates of the spin operator in
the direction of the  $\vec B$ field: $\hat s_B = \vec
B\cdot\vec\sigma$.   This is easy to see, and we
will prove it, for completeness, in the Appendix.  Thus, for large
$\vec B$ field, the modified  Stern--Gerlach apparatus is simply an
apparatus that measures the component of the spin along the
direction of the strong homogeneous magnetic field, splits
the beam and collapses the wave function accordingly.  Therefore,
taken literally, the claim in pg.9 of Ref.[1] that the beam does not
split is not correct.

A possible origin of confusion is given by the following fact:  Let
$\vec d\ $ be the direction along which the beam is deflected, and
let
the beam deflection be proportional to the component of the
magnetic moment of the particles along the direction $\vec r$. In a
conventional Stern--Gerlach experiment these two directions
coincide. However, in the modified Stern--Gerlach experiment they
do not: the beam is deflected along a direction given by the external
product between the strong homogeneous magnetic field and the
field gradient, namely
$$
\vec d = \nabla\vec B \times \vec B,
$$
but the amount of the deflection (and thus the splitting) is
proportional to the component of the magnetic moment along $\vec
B$, namely
$$
\vec r = \vec B.
$$
Therefore the modified Stern--Gerlach experiment splits the beam
in eigenstates of $\sigma_{\vec r} =: \sigma_B =: \vec
B\cdot\vec\sigma$, even if the beam is deflected in a
direction different than $\vec r$. By analogy with the conventional
Stern--Gerlach
experiment, one could be tempted to mistakenly assume that the
modified Stern--Gerlach experiment measures the component of the
particle's spin along the $\vec d$ direction; then the surprising
result that a beam polarized in the $\vec r$ direction does not split
follows. This result is precisely what is proven in Sec. 5 of Ref.[1],
where the claim that the beam does not split is made.  But the
modified Stern--Gerlach experiment does not measures the
component of the particle's spin along the $\vec d$ direction. It
measures the component along the $\vec r$ direction, therefore an
$\vec r$-polarized beam does not split simply because it is in an
eigenstate of the operator being measured. An arbitrarily polarized
beam will split.

\vfil\eject

A careful reading of the article, however, shows that the precise
setting of the experiment proposed by Aharonov and Anandan is more
subtle than just a Stern--Gerlach experiment with an additional
magnetic field.   In fact, the following assumption is made in the all
explicit the calculations of Sec.[5] of Ref.[1], as well as more
or less explicitly stated in several parts of the paper (for instance,
pg 10):

\vskip .4cm

{\it Assumption AA:\  } The initial state of the particle is
polarized in a direction $\vec r$ and this direction is
{\it the same\ } as the  direction of the homogeneous magnetic field
$\vec B$.

\vskip .4cm

This assumption is necessary for Aharonov and Anandan, since,
if we do not assume it, the beam splits, and therefore there is no
"protected measurement", as claimed. In other words, without
Assumption AA, the prediction made in  Ref.[1] on the fact that in
the modified Stern--Gerlach there is no beam splitting would be
wrong.  Thus, in the rest of the paper we discuss an experiment made
under the Assumption AA.

The immediate naive criticisms to an experiment made under
Assumption AA is the following: if we prepared a strong magnetic
field in the direction of the initial polarization of the particle, then
we already knew the initial polarization of the particle (perhaps up
to an overall sign). But the
initial polarization of the particle is what the experiment is
supposed to determine: if we know it, we know the wave function
up to a phase.  Therefore the experiment is not measuring anything.
It is just a complicated way of performing the following operation:
Given a particle with initial known spin wave function $\Psi$, have
it go through some dynamics that does not change $\Psi$. For
instance, a beam emerging from a conventional Stern--Gerlach
apparatus that measures the spin in the $z$ direction is polarized in
the $z$ direction; if we let it go through a {\it second\ }
conventional
Stern-Gerlach apparatus that again measures the spin in the $z$
direction, then the wave will not undergo further collapse.  However,
the second Stern--Gerlach apparatus is not a way of "measuring" the
wave function without collapse, since it the collapse is avoided
using the fact the polarization is already known.

Aharonov and Anandan, however,  go around this simple
criticisms by proposing the following setting for the experiment:
Not only we do not know the initial state; but also {\it we do not
know the orientation of the strong magnetic field $\vec B$.\ }   This
is stated repeatedly in Ref.[1] (for instance the beginning of pg10).
Indeed, what is actually measured by the experiment is the direction
of the strong additional magnetic field $\vec B$ (as explicitely
claimed, for instance, at the end of Sec. 5A), and the initial
polarization is inferred by the
fact that, thanks to Assumption AA, the initial polarization is
parallel to this direction.

Let us therefore summarize the experimental setting proposed in
Ref.[1]. The observer does not know the initial polarization
$\vec r$ of a single quantum particle, and does not know the
direction of a strong homogeneous magnetic field $\vec B$.
However, he does know that $\vec r$ and $\vec B$ are parallel
(and oriented in the same versus).  The observer then lets the
particle interact with the field  $\vec B$ and with a standard weak
inhomogeneous Stern--Gerlach magnetic field $\vec\beta$ over
which
he has complete control.  The result of this interaction is such that
the spin of the particle is not modified, but the observer  has
learned its initial polarization.  The claim is then made that this is
a way for measuring the polarization of the particle spin without
disturbing it.

We wish to argue in this note that this claim is unsubstantial.  The
main observation is that the strong homogeneous magnetic field
$\vec B$ is a macroscopic quantity, and is treated by Aharonov and
Anandan as a classical field.    Now, the Assumption AA
given above, requires that the field $\vec B$ is in the same direction
as the particle polarization.  Thus, the Assumption AA requires that
a macroscopic classical quantity has been correlated with the
particle polarization.  However there is no way of achieving this,
{\it but\ } by having already measured the particle polarization and
therefore having already disturbed and collapsed the particle wave
function.

To make the problem particularly evident, consider the following
experimental arrangement, which is entirely equivalent to the
Aharonov--Anandan experiment as far as the interpretation of
quantum mechanics is
concerned.  First measure the polarization of a quantum particle.
Second, write the outcome of this measurement on a piece
of paper.  At this stage assume that we do not know the
polarization and we do not know what is written in the piece of
paper.
Make then the following protected measurement: read what is
written on the piece of paper.  During the reading, the state of the
particle is not collapsed (we assumed it was already collapsed
during the first measurement), thus, the state of the wave function
is not affected by the reading.  During the reading we learn
about the state, thus we can detect the wave function without
disturbing it.  This is a very plausible description of a sequence of
events, but it is clearly meaningless as far as providing a new
interpretation of the wave function.

The analogy with the Aharonov--Anandan experiment is as follows.
The piece of paper is the analog of the strong magnetic field $\vec
B$, which is treated classically, contains a record of the quantum
state of the particle, and represents  what is actually measured
(read) in
the experiment.   The Aharonov--Anandan
experiment is obscured by the fact that one makes a classical
measurement of the direction of
$\vec B$, by using the particle itself; this fact simply obscures the
reading of the experiment.

To get more clarity, one should consider the following distinction. In
the Aharonov--Anandan experiment, and in
the limit of strong $\vec B$ field, the trajectory of the particle is
deflected.  Consider this deflection. Now, one should distinguish the
dependence of this deflection on the magnetic field from the
dependence on the particle's polarization.  For spin 1/2 particle this
is particularly simple: The direction toward which
the particle is deflected (the direction of the force), as well as the
absolute value of the deflection depend solely on the field, while the
sign (the versus) of the deflection depend on the polarization.
In the conventional Stern--Gerlach experiment
the direction of deflection is given by the
inhomogeneous magnetic field. Suppose that in a conventional
Stern--Gerlach experiment we do not know the direction of the
gradient of the field: we may consistently read it out from the
direction of the beam's deflection.  This is of course not a quantum
measurement: is a classical measurement of a classical macroscopic
quantity: the direction of the field gradient.
Thus, the deflection
of the beam contains two independent informations: one concerning
the orientation of the field, the other concerning the spin of the
particle. The first one is a classical "measurement" of a
macroscopic quantity (the orientation of the magnetic field), there
is no quantization and is fully deterministic; the second one is the
quantum measurement (of the spin of the particle), there is
indeterminacy and wave function
collapse.

In the modified Stern--Gerlach experiment, and in the large $\vec B$
limit, the direction of the deflection is the direction of the {\it
homogeneous\ } $\vec B$ field, and its absolute vale depend solely
on the fields.  On the other side, the sign of the deflection depends
on
the initial polarization of the particle.  Thus, in the presence of the
strong $\vec B$ field,
we can have a simple classical determination of the direction of
$\vec B$ by looking at the direction of the beam deflection: this is a
classical measurement of a macroscopic quantity, the outcome is a
continuous number, and it does not imply any collapse.   On the other
side, the fact that the deflection is in one versus or in the opposite
one, depends on the initial
polarization.  For an arbitrary initial polarization, there is wave
function collapse, no way to predict the outcome with certitude and
so on.

In the Aharonov--Anandan experiment, the particle's wave function
does not collapse simply because it is in an eigenstate of the
operator that it is being measured by the apparatus (the spin along
the $\vec B$ field).   Thus, as far as the particle's polarization is
concerned, the situation is fully analogous to a particle going
through a sequence of conventional Stern-Gerlach apparata all
oriented in the same way: no collapse happens.  However, the
direction of the deflection is used by  Aharonov and Anandan as a
way to make a (classical) measurement of the direction of $\vec B$.

It is the fact that these two "measurements", which must be kept
conceptually well separated (the classical measurement of the
direction of  $\vec B$, and the quantum measurement of the
polarization) are performed  simultaneously, which obscures the
interpretation of the Aharonov--Anandan experiment.  Once this has
been disentangled, the interpretation of the experiment is simple:

i.  The particle does not collapse because it goes through an
apparatus that measures an operator of which it is in an eigenstate.

ii.  The deflection of the trajectory is a classical measurement of a
macroscopic classical quantity, namely the orientation of $\vec B$.

iii. The experiment provides information about the spin polarization
only because it was {\it assumed\ } that the macroscopic quantity
$\vec B$ contains information about the polarization.

iv. The reason for which it seems that we can determine the
polarization without collapsing the wave function, is simply because
we assumed that the measurement of the polarization {\it had
already happened before,\ } and the outcome was stored in a
macroscopic object treated classically ($\vec B$: the "piece of
paper").

In other words, once we accept Assumption AA, we have already
measured the spin of the particle, and already made the wave
particle collapse.   Thus, the core of our argument is the following:
Given a particle in an {\it unknown\ } polarization state, {\it there
is no way of constructing a macroscopic magnetic field parallel to
the particle polarization, without disturbing the initial polarization
of the particle.}

\vskip1cm

Of course, the experiment can be analyzed in a variety of alternative
ways. For instance one can assume that the wave function did not
collapse in the interaction that correlated it with $\vec B$, and thus
$\vec B$ is in a quantum superpositions of macroscopically distinct
states. Then the magnetic field wave function would collapse during
the
modified Stern--Gerlach experiment, contrary to the no--collapse
claim.

Alternatively, or one can take any other interpretation
of quantum mechanics in which the wave function does not collapse
at all.  In an Everet--like interpretation, there would be no collapse,
but
there would be two distinct branches of the wave function, with
entanglement between the spin state and the center of mass
position; in a hidden variable theory the center of mass position will
depend on the hidden variables, and so on.  We do not see in any of
these interpretations any sense in which the Aharonov--Anandan
experiment should have a meaning substantially
different than the determination of the polarization of the particle
by reading the paper on which this spin was previously recorded.

It is well known in quantum mechanics, that if we assume that there
exists a single object that behaves non-quantum mechanically once,
then we could simultaneously measure non-commuting observables
of any other system interacting with it. This is the well known
argument at the roots of the thesis that every system must behave
quantum mechanically if the electron does.  In other words, if one
is allowed to cheat just once in quantum mechanics, then one can
disprove all the quantum mechanical standard results.  Assumption
AA is essentially the assumption that we already have gone once
around the basic fact that the measurement disturbs the system.

Finally, in order to show that the addition of the strong constant
magnetic field is not at all related to the results claimed in Ref.[1],
consider the following modified version of the Aharonov--Anandan
experiment.  Let us consider a particle moving along the $y$
direction, polarized in a direction $\vec p$ unknown to us. Let the
particle go through a {\it completely conventional\ } Stern-Gerlach
inhomogeneous magnetic field $\vec B_{conv}$, oriented in such a
way that it would split the two eigenstates of $\sigma_p= \vec
p\cdot\vec\sigma$ (thus, to orient it we have to know the
polarization already).   However, let us assume that the orientation
of this  $\vec B_{conv}$ field is {\it unknown to the observer}, in the
same sense in which Aharonov--Anandan assume that the strong
homogeneous field is unknown in their experiment. Finally, let us
imagine that the observer looks at the deflection of the particle's
trajectory from the straight $y$ direction.   Now, the following
holds:  i. from the trajectory's deflection, the original polarization
can be entirely reconstructed;  ii. the state does not collapse.
Therefore, according to the  Aharonov--Anandan definition, this is a
protected measurement.  This example shows that the "protection" is
given by the fact that the quantum system starts in an eigenstate of
the operator being measured.  The addition of the strong magnetic
field has the only effect of obscuring the physics, and has no
relevant "protective" effect whatsoever.

In conclusion, we claim the following:  If the Aharonov--Anandan
experiment is performed with a fixed $\vec B$ field, and an
arbitrary initial polarization, then the beam splits, contrary to the
claim of Ref.[1]. If, on the other side, Assumption AA is made, (a
classical $\vec B$ is parallel to the initial polarization), then the
experiment does not bear at all on the issue of the possibility of
measuring a system without disturbing it, because Assumption AA
means the the polarization had been already measured (and therefore
the wave function collapsed) in the past.  The "strong magnetic field
does not protect anything, it just changes the quantity being
measured.  A "protected  measurement" turns out to be nothing but a
very conventional  measurement plus the assumption that prior to
the measurement the system is already in an eigenstate of the
operator being measured.  A situation in which, to nobody's surprise,
no collapse occurs.

\vskip 1cm

I thank  Jim Bayfield, Al Janis and Ted Newman for the stimulating
discussions on the subject.
\vskip .5cm

Pittsburgh, April 21, 1993.

\vskip 2cm

\bf Appendix \rm

\vskip .5cm

In this appendix we show that for a fixed $\vec B$, and in the large
$\vec B$ limit, a beam with arbitrary polarization splits in the
eigenstates of the spin along the $\vec B$ direction.

This can be done just by repeating the calculations of Sec. 5 of Ref.
[1], but with the opposite, orthogonal, initial state. The final
deflection, or the final time delay in the case considered in Sec. 5A,
turn out to have opposite sign. Therefore a beam with arbitrary
initial polarization will split in the two components polarized in the
direction of the strong field.

However, there is a simpler way to get to the same
result, which may better illuminate the physics of the experiment.
Here
we will discuss this alternative derivation. If the present derivation
is found unsatisfactory, one can resort to the fully quantum
mechanical derivation, and confirm the results obtained here.

We begin by providing a classical description of the experiment. The
classical dynamics of a particle with magnetic moment $\vec M$,
flying through a magnetic field $\vec B$, is easy to work out. The
center of mass of the particle feels a force proportional to the
component of $\vec M$ normal to the gradient $\nabla\vec B$ of the
magnetic field, and $\vec M$ precesses around the magnetic field
$\vec B$ at the Larmor frequency.   Assuming the homogeneous
component of the magnetic field to be very
strong compared with its gradient and with the flying time,  $\vec
M$ will precess very rapidly around $\vec B$, so that the force felt
by the center of mass is partially averaged out to zero.  More
precisely, the force on the particle's
center of mass due to the component of $\vec M$  normal to  $\vec
B$ is averaged out to zero by the Larmor precession.  The force that
is not averaged out is the one due to the component of $\vec M$
parallel to  $\vec B$, namely to
$$
\vec M_B = {\vec M \cdot \vec B \over |B|^{2}}\ \vec B =  M_B\
{ \vec B\over  |B|}.
$$
We have introduced the component $M_B$ of $\vec M$ along $\vec B$,
as
$$
	M_B ={\vec M \cdot \vec B \over |B|}.
$$
The average force felt by the center of mass is therefore given by
$$
\vec F = k\   \nabla\vec B \times \vec M_B
= k\ 	{\nabla\vec B \times \vec B \over |B|}\ M_B .
$$
where k is a constant.  The integration of the trajectory is
immediate, and we obtain that the particle is deflected as follows.
The trajectory is deflected in the {\it direction\ } of the vector
$$
\vec d = \nabla\vec B \times \vec B ,
$$
by an amount $D$ given by
$$
D  =  c\  {|(\nabla\vec B \times \vec B ) | \over |B|}\ M_B
$$
where c is a constant that depends on the initial speed, the space
extension of the field, and so on. In other words:
$$
  D =  C \ M_B,    \eqno(1)
$$
where $C$ depends on $\nabla\vec B$ and $ \vec B$ but does not
depend on the magnetic
moment. The deflection $D$ depends on the initial polarization of the
particle, namely
on its initial magnetic moment $\vec M$, only through $M_B$,
namely through the component of  $\vec M$ along
$\vec B$.

Let us now consider the quantum theory. As in Ref.[1], we take the
observed  "system" to be the
particle's spin direction. The separation between the
system and the
apparatus is of course conventional and does not affect the
predictions,
as far as there is some classical apparatus. The difference between
the treatment here and the one in Ref.[1] is that we treat the center
of mass classically, while in Ref.[1] it is treated as part of the
apparatus, but still quantum
mechanically. This does not affect the final result in any way.  The
relevant Hilbert
space is the two dimensional Hilbert space of the particle's spin.
The key point is to understand which
is the operator that describes the measurement. The number
that represents the outcome of the (classical) measurement is the
displacement $\vec D=D\vec d= C\vec d M_B$. Here $C$ and $\vec d$
depend on classical quantities. Indeed, $\nabla\vec B$
and $\vec B$ are treated as classical fixed external fields: for any
given experimental
configuration the quantity  $C\vec d$  is a c-number, that depends on
the
apparatus. Therefore, as far as the quantum system is concerned,
{\it the experiment measures $M_B$, that is, the component of the
magnetic moment of the particle in the direction of the magnetic
field.\ }   The operator that represents the observable that
is being measured is the component  $\hat M_B= (\hat{\vec M}\cdot
\vec B) |B|^{-1})$  of the magnetic moment in the direction of the
field $\vec B$, where $\hat{\vec M}=\gamma \hat{\vec S}$ is the
magnetic moment operator and $\hat{\vec S}$ is the spin operator.
The possible outcomes of the experiment are given
by the eigenvalues of  $\hat M_B$.    These are of course given by
plus or minus one half the Planck constant times the spin--magnetic
ratio $\gamma$ of the particle. So we conclude that the possible
deflections are
$$
\vec D_+= + {\hbar\over 2}\  \gamma\ C \vec d ,\ \ \ \
\
\vec D_- = - {1\over 2}\ \hbar \gamma\ C \vec d.
$$
A generic state will be a quantum superposition of the two
eigenstates of $\hat M_B$, and the probability that the particle
will land in one or the other possible spot will be given by the
amplitude of the two respective components.  We stress the fact
that $C$ and $\vec d$ depend on $\nabla\vec B$ and $\vec B$, as
indicated.   Thus, the position of
the spots where the particle may land depends continuously on the
field, and can be used as a classical measurement of the fields. Note
also that the direction of deflection, $\vec d$, is different than the
direction along which the magnetic moment is measured, $\vec B$ as
indicated above in this note.

The important point here is that there are {\it two\ }
eigenvalues of the operator measured, and therefore two possible
outcomes of the measurement, and not one, as indicated in [1]. For a
generic linear combination of eigenstate of the spin along $\vec B$,
the beam splits.

In the paper [1], the authors present an explicit calculation in which
they show that the splitting does occur for a weak added magnetic
field, but does not occur in the limit in which the added magnetic
field is strong.   The calculation in Ref. [1] is performed with a very
specific setting: namely {\it the initial state is always taken to be
an eigenstate of the magnetic field in the direction of the strong
homogeneous magnetic field\ } $\vec B$.  More precisely,  $\vec B$
is taken in the $z$ direction, and the initial state is always taken to
be a state $(1,0)$ in the basis that diagonalizes the spin operator in
the $z$ direction. As far as  $\vec B$ is not strong, the experiment
measures some component of the spin different than $z$, and
therefore there is splitting.  But in the strong  $\vec B$ limit the
spin component that is being measured is precisely the $z$ one,
thus, the  reason for which the authors do not get any splitting is
simply because they assume the particle is already in an eigenstate
of the operator being measured.

\vskip2cm
\bf References \rm
\vskip.3cm

[1] Y Aharonov and J Anandan, "Meaning of the wave function", to
appear on Phys Rev A (June1, 1993).

[2] David Freedman, "Theorists to the Quantum Mechanical Wave: 'Get
Real'\ ", Science Vol. 259, pg 1542 (1993).

\end{document}